\documentclass[]{revtex4}
\usepackage{amsmath}    
\usepackage{graphicx}   
\usepackage{verbatim}   
\usepackage{color}      
\usepackage{subfigure}  
\usepackage{hyperref}   
\usepackage{lineno} 
\usepackage{bm} 
\usepackage{multirow}
\usepackage{csquotes}
\usepackage{appendix}
\newcommand{\classoption}[1]{\texttt{#1}}

\expandafter\ifx\csname package@font\endcsname\relax\else
\expandafter\expandafter
\expandafter\usepackage
\expandafter\expandafter
\expandafter{\csname package@font\endcsname}%
\fi
\DeclareRobustCommand\substyle{\name@idx{document substyle}}%
\DeclareRobustCommand\classoption{\name@idx{document class option}}%
\DeclareRobustCommand\classname{\name@idx{document class}}%
\def\name@idx#1#2{
	{\ttfamily#2}%
	\index{#2\space#1=\string\ttt{#2}\space#1}\index{#1>#2=\string\ttt{#2}}%
}
\begin{document}

	\pagenumbering{arabic}
	\title{Study of pion production in $\nu_{\mu}$ interactions on $^{40}$Ar in DUNE using GENIE and NuWro event generators }
\author{ H.R. Sharma$^{1}$\footnote{E-mail: hansraj77sharma@gmail.com}, Srishti Nagu$^2$\footnote{E-mail: srishtinagu19@gmail.com}, Jyotsna Singh$^2$\footnote{E-mail: singh.jyotsnalu@gmail.com }, R.B. Singh$^2$\footnote{E-mail: rajendrasinghrb@gmail.com }, Baba Potukuchi$^1$\footnote{E-mail: baba.potukuchi@gmail.com} }
\affiliation{Department of Physics, University of Jammu, Jammu, India$^{1}$}
\affiliation{Department of Physics, Lucknow University, Lucknow, India$^{2}$}

	\begin{abstract}
 The study of pion production and the effects of final state interactions (FSI) are important for data analysis in all neutrino experiments. For energies at which current neutrino experiments are being operated, a significant contribution to pion production is made by resonance production process. After its production, if a pion is absorbed in the nuclear matter, the event may become indistinguishable from quasi-elastic scattering process and acts as a background.  The estimation of this background is very essential for oscillation experiments and requires good theoretical models for both pion production at  primary vertex and after FSI. Due to FSI, the number of final state pions is significantly different from the number produced at primary vertex. As the neutrino detectors can observe only the final state particles, the correct information about the particles produced at the primary vertex is overshadowed by FSI. To overcome this difficulty, a good knowledge of FSI is required which may be provided by theoretical models incorporated in Monte Carlo (MC) neutrino event generators. They  provide theoretical predictions of neutrino interactions for different experiments and serve as a bridge between theoretical models and experimental measurements. In this work, we will present simulated events for two different MC generators - GENIE and NuWro, for pion production in $\nu_{\mu}$CC interactions on $^{40}$Ar target in DUNE experimental set up. A brief outline of theoretical models used by generators is presented. The results of pion production are presented in the form of tables showing the occupancy of primary and final state pion topologies with 100$\%$ detector resolution and with kinetic energy detector threshold cuts. We observe that NuWro (v-19.02.2) is more transparent (less responsive) to absorption and charge exchange processes as compared to GENIE (v-3.00.06), pions are more likely to be absorbed than created during their intranuclear transport and there is need to improve detector technology to improve the detector threshold for better results.\\
	\end{abstract}
	\maketitle

\section{Introduction}
\label{section1}
Driven by new experiments and modern detector technology, neutrino physics is entering a precision era and this requires an improved theoretical and phenomenological description of neutrino interactions. Neutrinos rarely interact with matter and can travel long distances without any interactions. The neutrino properties are still not entirely understood and this makes its research a challenge for both theoretical and experimental points of view. Electroweak theory of Standard Model (SM) describes the neutrino interactions. In its earlier formulation, SM assumed the neutrino to be massless particle and and so the mass-mixing was not expected unlike quarks. The mass mixing would also be possible in lepton sector if neutrino had mass and a neutrino produced in a specific flavour could be later seen as neutrino of another flavour, the phenomenon called neutrino oscillations. No fundamental principle in physics requires neutrino to be massless. The strength of mass mixing in leptonic sector is defined by Pontecorvo-Maki-Nakagawa-Sakata (PMNS) matrix\cite{Giganti:2017fhf}. PMNS can be expressed by three mixing angles ($\theta_{12}$, $\theta_{23}$, $\theta_{13}$) and CP phase factor ($\delta_{cp}$). Apart from these mass mixing parameters, the probability for neutrino oscillation depends on actual neutrino masses (or on the difference of their squares) i.e. $\Delta m^2_{21}$ = $m^2_2$ - $m^2_1$; $\Delta m^2_{32}$ = $m^2_3$ - $m^2_2$; $\Delta m^2_{31}$ = $\Delta m^2_{32}$ + $\Delta m^2_{21}$.\\

For current and future neutrino experiments, understanding of charged current (CC) neutrino-nucleus interactions in few GeV energy region is very important. But the study of these interactions in this energy region is complicated and requires many intermediate steps, such as understanding the neutrino-nucleus cross-sections, description of nuclear models, modelling of hadronization and intranuclear hadron transport and nuclear effects. For this we require a canonical Monte Carlo generator which may take into consideration all these steps. There are a number of Monte Carlo generators dedicated to the description of neutrino interactions like ANIS \cite{Gazizov:2004va}, GENIE \cite{Andreopoulos:2015wxa}, NuWro \citep{Golan:2012wx,Sobczyk:2008zz}, GiBUU \citep{Buss:2011mx,Gibuu}, MARLEY \cite{Gardiner:2021qfr}, FLUKA \cite{Battistoni:2009zzb}, NEUT \cite{Hayato} and Nuance \cite{Casper:2002sd}. All these generators are based on similar assumptions. The primary neutrino interactions and final state interactions are considered separately in each generator.\\

In this work, GENIE and NuWro generators are taken into consideration for simulation work of pion production in neutrino-nucleus interactions. Pions form an important background \cite{MicroBooNE:2021sne} in many oscillation experiments and are also theoretically challenging due to processes they undergo in FSI \cite{Naaz:2018amr}. For both GENIE and NuWro, DUNE flux has been used for interactions on $^{40}$Ar target. For each generator only the charged current (CC) interactions were considered. The processes enabled were quasi-elastic (QE) scattering, resonance (RES) production, deep-inelastic scattering (DIS), coherent (COH) pion production. The data generated was then analysed for various pion topologies before and after final state interactions (secondary
interactions). It is crucial for the presently running experiments like T2K \cite{T2K:2019bcf} and NOvA \cite{NOvA:2019cyt} and the future long-baseline neutrino-oscillation experiments like DUNE \citep{DUNE:2020lwj,DUNE:2020ypp,DUNE:2021cuw} and Hyper-KamioKande \cite{Hyper-Kamiokande:2018ofw} to clearly understand neutrino-nucleus interactions.\\

The DUNE (Deep Underground Neutrino Experiment) is a worldwide effort to construct a long-baseline neutrino oscillation experiment at Fermi National Accelerator Laboratory (FNAL) USA. It has a Near Detector (ND) located 575 m from the neutrino source and 60 m underground \cite{DUNE:2020jqi} on the Fermilab site in Illinois and a Far Detector (FD) located approximately 1.5 km underground at a distance of nearly 1300 km from Fermilab at Sanford Underground Research facility (SURF) in South Dakota, USA. The primary scientific goals of this extremely advanced detector are to carry out a comprehensive program of neutrino oscillation measurements using $\nu_{\mu}$ and $\overline{\nu}_{\mu}$ beams from Fermilab and constraining the CP violation phase in leptonic sector. As the distance between FD and ND is about 1300 km, it will provide a baseline facility of length 1300 km to study matter effect. Both ND and FD will use Ar target material to observe the neutrino spectrum that will help to overcome various systematic uncertainties. ND will observe unoscillated neutrino spectrum and FD will observe oscillated neutrino spectrum. ND and FD at DUNE will have different dimensions and technology. The DUNE Near Detector has three primary detector components and two of those components have capability to move off the beam axis: 1) A 50 ton LArTPC (ND-LAr) constructed using ArgonCube with pixellated read out 2) The ND-GAr detector (also sometimes called the multipurpose detector, or MPD) that consists of high pressure gaseous argon TPC surrounded by an electromagnetic calorimeter (Ecal) in a 0.5 T magnetic field 3) An on-axis beam monitor called System for on-Axis Neutrino Detection (SAND) that monitors the flux of neutrinos. It consists of an inner tracker surrounded by an Ecal inside a large solenoidal magnet. In current scenario, two options are being explored for the inner tracker, one based on a combination of plastic scintillator cubes with TPCs and other based on straw-tubes. \\

The capacity of ND-LAr and ND-GAr to move to take data in positions off the beam axis is reffered to as DUNE-PRISM (DUNE Precision Reaction Independent Spectrum Measurement). The DUNE FD consists of four similar LArTPCs, each LArTPC will have fiducial mass of at least 40 kt. Each LArTPC detector will be installed in a cryostat with internal dimensions 15.1 m (w) $\times$ 14 m (h) $\times$ 62 m (l) with a total LAr mass of about 17.5 kt. Excellent tracking and calorimetry performance will be provided by LArTPC which makes it an ideal choice for the DUNE-FD. \\

Accelerator generated beams are used in many long baseline neutrino oscillation experiments\cite{Nagu:2019uco}. As these neutrino beams are not mono energetic, reconstruction of neutrino energy is required and for that complete information of final state particles is necessary. The identification of primary state particles in the presence of nuclear effects is a difficult task since the particles produced in the primary neutrino-nucleus vertex and the particles captured by the detector can be different because of FSI. This means energy reconstruction of neutrinos from final state particles needs careful examination. Neutrino experiments use heavy nuclear targets for large event statistics (which reduces statistical errors) but the use of heavy targets gives boost to nuclear effects thereby shifting the attention to systematic errors. The current knowledge of nuclear effects is still insufficient to have clear understanding of systematic errors\cite{Naaz:2018amr}. It must be emphasized that success of any neutrino experiment depends on successful understanding, quantifying and reducing of systematic errors coming from modelling neutrino-nucleus interactions. \\

\begin{figure}[h!]
 	\centering     
 	\includegraphics[scale=0.30]{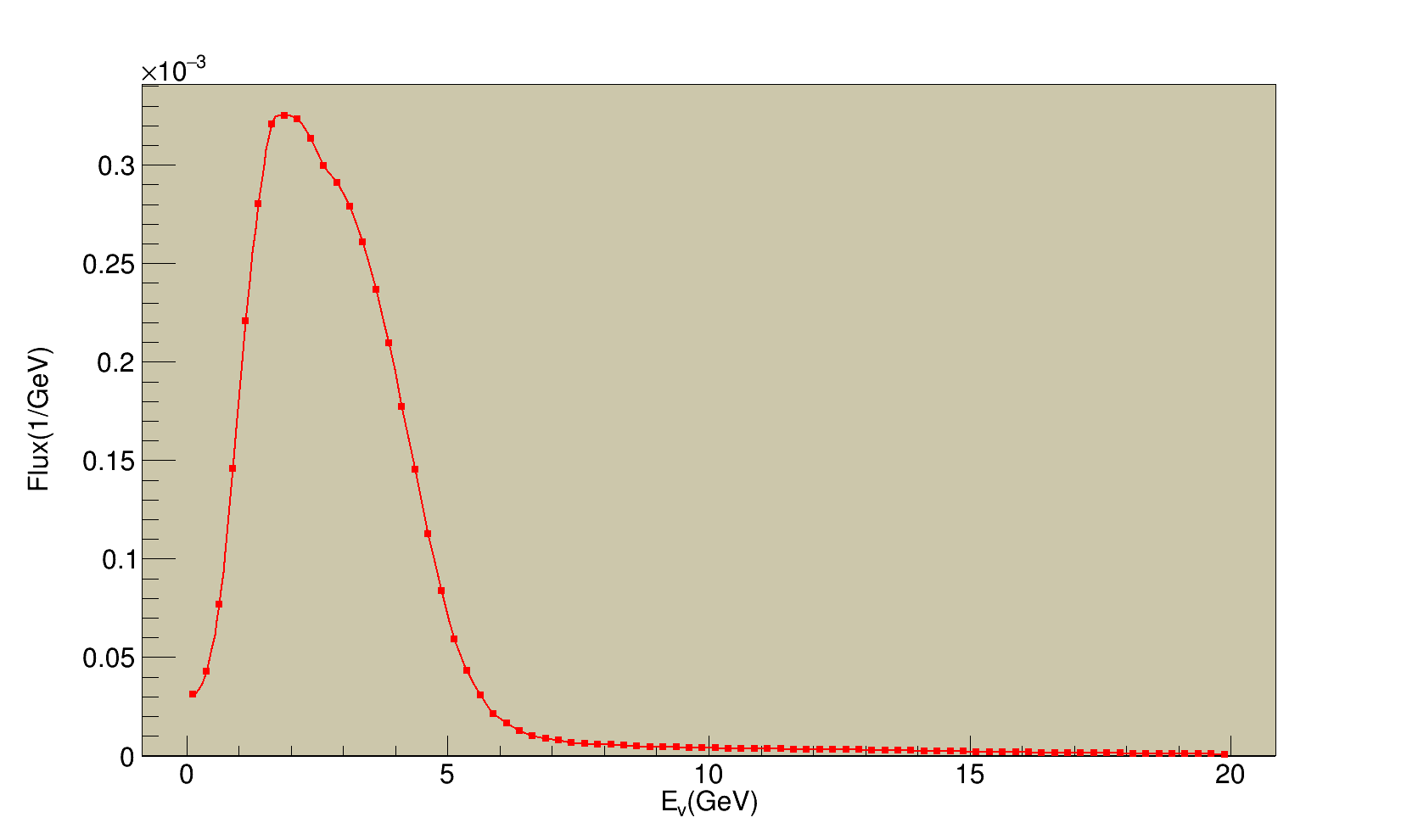}
 	
 	\caption{DUNE flux as a function of neutrino energy used in our work.}
 	\label{flux}
 \end{figure}

In our simulation, the DUNE neutrino flux \cite{flux:link} in the energy range 0.125-19.875 GeV on  $^{40}Ar$ nuclei was used. The neutrino flux used in our simulation work is shown in Fig. \ref{flux}. It covers the energy spectrum from about hundred  MeV to tens of GeV and peaks around 2.5 GeV. The Neutrino at Main Injector (NuMI) beamline facility at Fermilab provides an intense, high purity, wide-band neutrino beam with an initial power of 1.2 MW (will be further upgraded to 2.4 MW) for which 1.1$\times$10$^{21}$ protons are expected per year from the accelerator. The primary beam of protons coming from the main injector accelerator in the energy range 60-120 GeV is made to smash on the graphite target which results in the production of pions and kaons. These mesons will be further focussed toward a 200 m long decay pipe with the help of magnetic horns where they will decay into neutrinos and leptons. The neutrino and anti-neutrino beams can be separately ejected by changing the polarity of focussing magnets.\\

This paper is organised into following sections: Section-II contains description of pion production in neutrino-nucleus interactions. In this section QE, RES, DIS and COH processes are discussed in detail. An outline of GENIE and NuWro MC generators with various models used by them is given in Section-III. Results of simulation are given in Section-IV followed by summary and conclusions in Section-V.\\

\section{Pion Production in neutrino-nucleus interactions}
\label{section2}
The simplest description of neutrino-nucleus interactions is built upon two attributes: a characterization of neutrino-nucleus scattering and a model for nucleons in the nucleus. The modelling of neutrino-nucleus interactions is complex as it requires linking together many different pieces of theory. We will focus on neutrino-nucleus interactions in the nuclear environment. The description of hadronization process and final state interaction models are often specific to a particular generator.\\
In its simplest and most common approach (called impulse approximation approach), neutrino-nucleus scattering is treated as the incoherent sum of scattering from free nucleons in the nucleus. But practically, nucleons in the nucleus are not independent particles but bound states and more complex nuclear dynamics are involved in calculating the cross-section. The total cross-section of neutrino-nucleus charged-current scattering has the form \cite{Kuzmin:2005bm}:
\begin{equation}
\sigma^{tot}_{\nu N}=\sigma^{QE}_{\nu N}+ \sigma^{1\pi}_{\nu N}+ \sigma^{2\pi}_{\nu N} + ........+ \sigma^{1K}_{\nu N} +...... +\sigma^{DIS}_{\nu N}
\label{equation1}
\end{equation}
Here $\nu$ represents neutrino, N is nucleon, $\sigma^{tot}_{\nu N}$ is sum total of all cross sections, $\sigma^{QE}_{\nu N}$ is cross section for QE scattering, $\sigma^{1\pi}_{\nu N}$ is cross section for single pion production and so on. \\

\begin{figure}[h!]
  	\centering     
  	\includegraphics[scale=0.30]{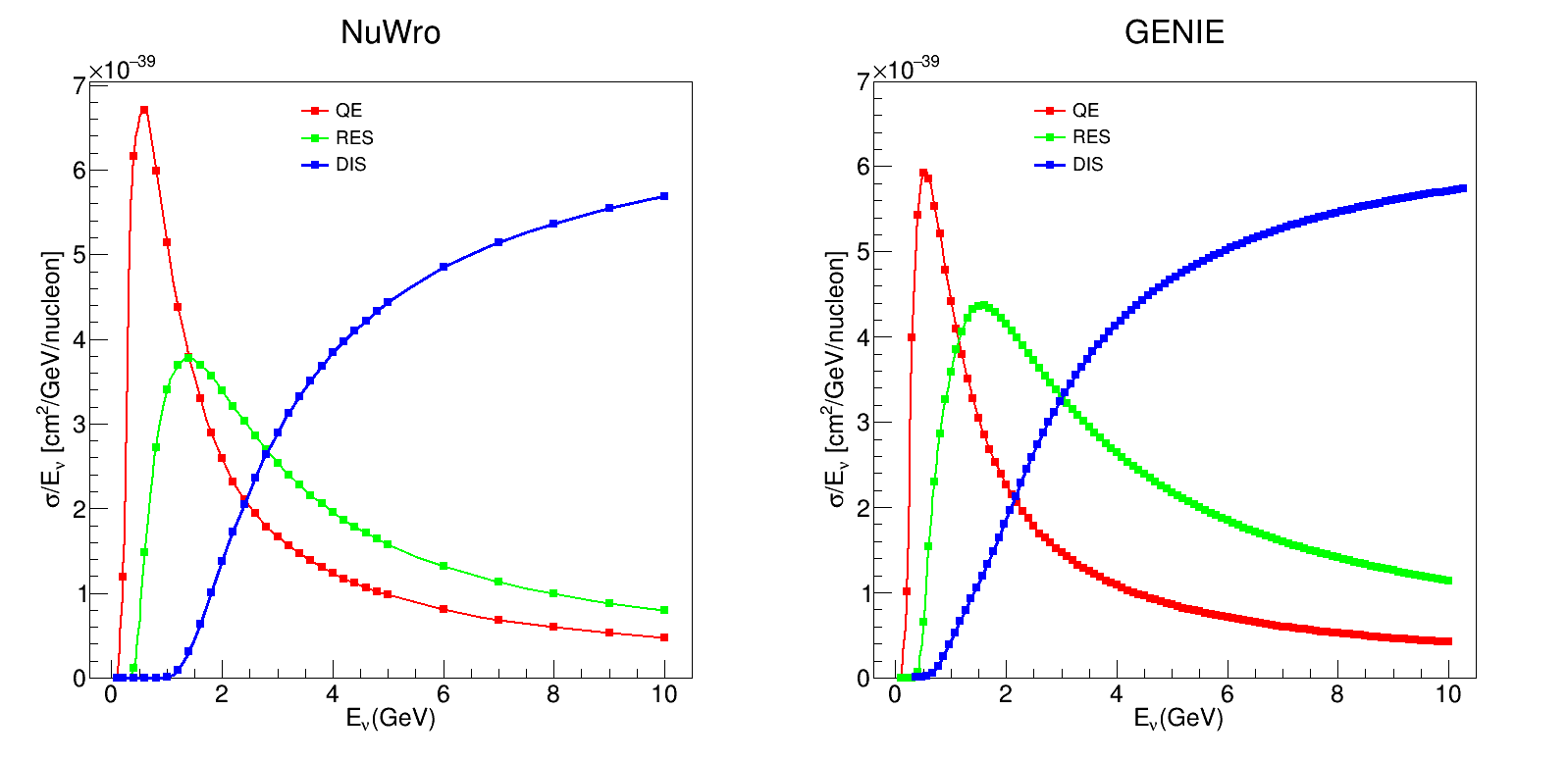}
  	
  	\caption{Neutrino-nucleus cross-sections on $^{40}Ar$ target for QE, RES and DIS processes, measured using NuWro (left) and GENIE (right)  generators.}
  	\label{xsec}
  \end{figure}

Neutrinos interact weakly with matter by the exchange of $W^{\pm}$ and $Z^0$ bosons. At low neutrino energies, QE scattering process is favoured. With the increase in neutrino energy RES and then DIS processes become predominant (Fig. \ref{xsec}). DUNE flux peaks around 2.5 GeV and at this energy  RES and DIS cross-sections have almost similar magnitudes. This is a problem experimentally, as in a detector, RES events can have signatures indistinguishable to DIS events. Thus, it is hard to measure each process separately.\\ 

 In QE scattering target nucleon remains a single nucleon in the final state and only changes its charge in CC weak interactions. In this scattering no pions are produced directly, but can be produced through final state interactions. Inside the nuclear environment, hadrons can be scattered elastically or inelastically, can be absorbed or charged exchanged and can even produce more pions. Thus, it is expected that a small number of events with no pions in primary state may produce pions in the final state. For $\nu_{\mu}$ beam, the CC QE scattering reaction is written as:
 \begin{equation}
 \nu_{\mu} + n \longrightarrow \mu^{-} + p
 \label{equation2}
 \end{equation}

  Here $\nu_{\mu}$ is muon neutrino, n is neutron, $\mu^{-}$ is muon and p is proton.\\

\begin{figure}[h!]
 	\centering     
 	\includegraphics[scale=0.50]{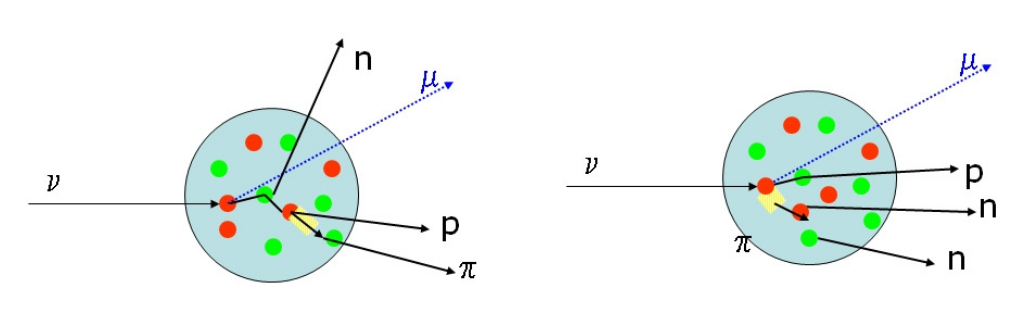}
 	
 	\caption{Pion production in QE scattering (left) and absorption in RES scattering (right) \cite{Andreopoulos:2015wxa}}
 	\label{pion_production}
 \end{figure}

\begin{figure}[h!]
	\centering     
	\includegraphics[scale=0.50]{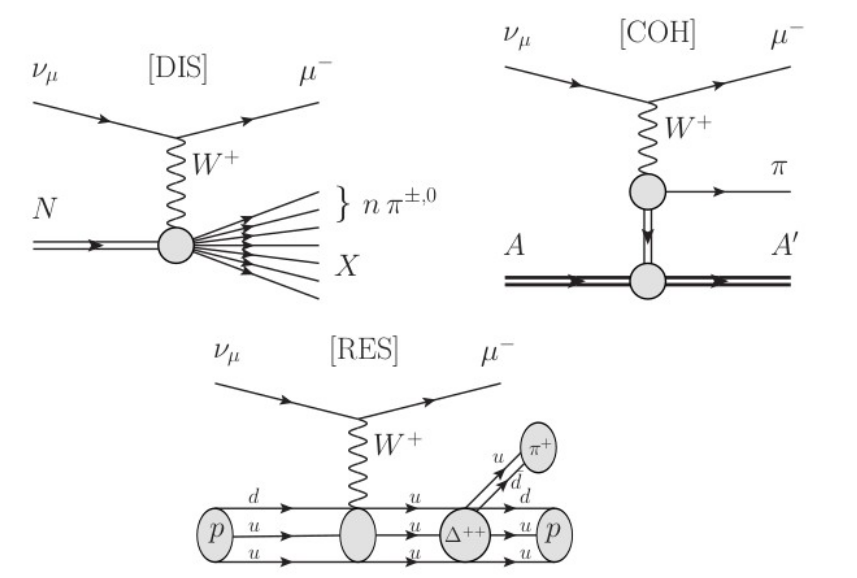}
	
	\caption{Pion production via various processes\cite{Antonello:2009ca} }
	\label{processes}
\end{figure}

  Fig. \ref{pion_production} (left) shows how pions are produced in QE scattering process and  how a pion can be absorbed (right) inside the nucleus \cite{Andreopoulos:2015wxa} during its intranuclear transport. Although the dominant processes that produce pions directly are DIS, RES and COH. These processes are shown in Fig. \ref{processes} \cite{Antonello:2009ca}. 
 
In RES scattering, pions are produced from resonances. In RES production process, a neutrino excites the target nucleon to a resonance state. The produced resonance state quickly decays into a nucleon and a single pion state. The dominant contribution from this process comes from $\Delta$ (1232) resonance state, although production of higher resonance states is also possible. The CC RES scattering processes are illustrated as under (for $\nu_{\mu}$ beam):

\begin{equation}
\nu_{\mu} + p \longrightarrow \mu^{-} + \Delta^{++}; \Delta^{++} \longrightarrow p + \pi^+
\label{equation3}
\end{equation}

\begin{equation}
\nu_{\mu} + n \longrightarrow \mu^{-} + \Delta^{+}; \Delta^{+} \longrightarrow n + \pi^+
\label{equation4}
\end{equation}

\begin{equation}
\nu_{\mu} + n \longrightarrow \mu^{-} + \Delta^{+}; \Delta^{+} \longrightarrow p + \pi^0
\label{equation5}
\end{equation}

If a pion produced in a RES process gets absorbed in the nucleus then it becomes difficult to identify whether this process is a RES process. Thus the resulting event gives the impression of a different process and is called a fake event. In a particular interaction channel, pure events are those in which particles detected by the detector are the same as were produced at the primary vertex.\\

In DIS, a high energy neutrino penetrates deep inside the nucleon and scatters off a quark in the nucleon via exchange of W and Z bosons producing a lepton and a hadronic system in the final state. For CC $\nu_{\mu}$ interaction, the process looks like:

 \begin{equation}
 \nu_{\mu} + N \longrightarrow \mu^{-} + n \pi^{\pm} + X
 \label{equation6}
 \end{equation}
 Where N is a nucleon (proton or neutron), n is a number and X is any set of final nucleons.\\
 
Neutrinos can also interact with the whole nucleus coherently producing pions (coherent (COH) pion production). For CC $\nu_{\mu}$ interaction, the process can be written as:\\

\begin{equation}
\nu_{\mu} + A \longrightarrow \mu^{-} + A' + m^+
\label{equation7}
\end{equation}
Where A is the nucleus in initial state, $A'$ is the nucleus in final state and $m^+$ = $\pi^+$, $k^+$, $\rho^+$ .......\\

The common approach with Monte Carlo generators is to use the Relativistic Fermi Gas (RFG) model \cite{Smith2} modified by Bodek and Ritchi to describe the nuclear environment. In the model, fermi motion of individual nucleons is taken into account but its implementation often differs for different neutrino generators. The importance of considering Pauli blocking \cite{Benhar:2006nr}, impulse approximation \cite{Frullani:1984nn} and use of spectral functions \cite{Benhar} is also taken into account for better description. The description of hadronisation and the propagation of secondary particles to come out of the nucleus is also necessary. The simulation must cover the description of both re-scattering and absorption effects and for this modelling of final state interactions is required.\\

\section{GENIE and NuWro Monte Carlo Generators}
\label{section3}
Neutrino event generators are the simulation tools which are used in the study of  neutrino physics and these generators can be optimized using experimental data gathered from some previous experiments. Generators work as a bridge between experimental and theoretical frameworks. The two neutrino event generators used in this work are GENIE (Generates Events for Neutrino Interaction Experiments) and NuWro (developed by Wroclaw University). In our simulation studies, we have used version 3.00.06 of GENIE and version 19.02.2 of NuWro which are the latest stable releases. GENIE is a universal neutrino event generator, developed by international collaboration and written in C++. It is a modern and most sophisticated platform for simulating neutrino events and is developed by taking into consideration ongoing neutrino oscillation experiments. It simulates neutrino interactions for all neutrino flavours and all targets over energy range from few MeV to several hundred GeV. This generator has already been used by experiments like MINOS \cite{Evans:2013pka}, MINERvA\cite{MINERvA:2004gta}, T2K\cite{T2K:2019bcf} , MicroBooNE\cite{MicroBooNE:2007ivj}, NOvA\cite{NOvA:2019cyt} and AgroNEUT\cite{Soderberg:2009qt}. NuWro is more theory oriented, developed by Wroclaw University and written in C++. It is simple but complete package for generating neutrino-nucleus interactions. NuWro clearly specifies
incoming (in), primary state (out) and final state (post) particles of an interaction. In this generator simulation parameters can be fixed in a plane text file (params.txt) and it gives output in a root file.  For scattering off free nucleon, NuWro can simulate events for neutrino energies from threshold to TeV. In NuWro, a beam can be set manually or can be loaded from ROOT format file, there are choices for description of target nucleus including Fermi gas and spectral functions, there are various parametrizations of nuclear form factors and there is a detector geometry module. Many quantum effects like formation zone and Pauli blocking are included. This makes NuWro a good tool for the use in neutrino experiments. Recently it has been successfully used by experiments like T2K and MINERvA.

\subsection{GENIE}
The physics models used in GENIE can be easily understood by classifying them into three categories: cross-section models, hadronization models and nuclear physics models. For cross-sections, charged-current quasi-elastic scattering is described using Llewellyn-Smith model \cite{LlewellynSmith:1971uhs} using latest BBBA07 form factors \cite{Bodek:2007wz}. Rein-Sehgal model\cite{Rein:1980wg} has been used for the production of baryon resonances. For DIS interactions, Bodek and Yang model \cite{Bodek:2002ps}, with modification at low $Q^{2}$, is used. Rein-Sehgal model with an updated PCAC formula \cite{Dieter} is used for coherent pion production interactions. The hadronization process employs default AGKY model \cite{GENIE:2021wox}. It includes a phenomenological description of low invariant mass region using Koba-Nielson-Olsen (KNO) scaling\cite{Koba:1972ng}, while at higher masses it gradually switches over to the PYTHIA/JETSET model, the transition being gradual and continuous. To take into consideration the effect of nuclear environment, implementation of Fermi Gas model with modification by Bodek and Ritchie, is used to include nucleon-nucleon correlations. The factors like Pauli blocking and differences between free nucleon and nuclear structure functions is also taken into consideration. The INTRANUCLEAR/hA model is used to handle intranuclear hadron transport.

\subsection{NuWro}
In NuWro, QE events are simulated using Llewellyn-Smith model with latest BBBA05 form factors \cite{Bradford:2006yz}. In RES, only $\Delta$(1232) contribution is based on Rein-Sehgal model and for rest of resonances, Adler-Rarita-Schwinger model\cite{Graczyk:2009qm} has been used. Rein-Sehgal model is used for simulating coherent pion production interactions. This Rein-Sehgal model\cite{Lalit} is different from that used in RES processes.  The RES region is defined by a cut on invariant mass as W$<$ 1.4 GeV. NuWro uses Quark-Parton model \cite{Sjostrand:2006za} to describe DIS events. The DIS contribution is turned on when W$>$1.6 GeV. In the region 1.4 GeV$<$W$<$1.6 GeV, RES contribution is linearly turned off as the DIS contribution is turned on. NuWro uses its own hadronization model together with Bodek-Yang model. The effects of nuclear environment are taken into consideration using Relativistic Fermi Gas (RFG) model.\\

\section{Results}
\label{section4}
For both GENIE and NuWro, we have generated 2 Million similar sets of events using DUNE flux. The number of pions produced in both primary and final states were obtained on event by event basis for each generator. These simulated results are presented in the form of tables which show topologies of pion produced in primary and final states. A primary state contains topology of that pions which are produced in the primary neutrino-nucleus interactions. A final state contains topology of that pions which are produced after any of the secondary interactions (like intra-nuclear scattering or absorption) may have taken place. For fair comparison of generators, a description of models used by the generators and physics choices has already been given.\\
\begin{table}[h!]
	\centering
	\begin{tabular}{|c|c|c|c|c|c|c|c|c|c|c|c|c|c| }
		\hline
		& \multicolumn{12}{|c|}{Primary Hadronic States }\\
		\hline
		Final states  & $0\pi $&$ \pi^{0} $& $\pi^{+} $&$ \pi^{-} $&$ 2\pi^{0}  $&$ 2\pi^{+} $&$ 2\pi^{-} $&$ \pi^{0} \pi^{+}$ &$ \pi^{0}\pi^{-} $&$ \pi^{+}\pi^{-}$&$ \geq 3\pi$ & total\\
		\hline
		$	0\pi$ & 407305 & 44000&132839&580&343&900&0&2763&45&710&727&590212\\
		\hline
		$	\pi^{0}$& 4751 &151844&78184&449&4747&945&0&15006&122&645&3021&259744\\	
		\hline	
		$	\pi^{+}$&7238&13618&424563&482&141&6715&0&14061&11&4520&3227&474576\\
		\hline	
		$	\pi^{-}$&1361&13113&5557&1787&178&22&0&801&120&4502&1458&28899\\
		\hline
		$	2\pi^{0}$&45&6167&12943&136&14478&395&0&7310&71&358&5805&47708\\
		\hline
		$	2\pi^{+}$&32&729&6223&30&22&17805&0&3191&01&238&4836&33108\\
		\hline
		$	2\pi^{-}$&0&180&43&08&31&01&0&07&32&26&257&585\\
		\hline
		$	\pi^{0} \pi^{+}$&228&11314&22979&314&1251&3664&0&79596&41&2462&10997&132846\\
		\hline
		$	\pi^{0}\pi^{-}$&26&6021&3469&283&1369&25&0&928&467&2370&3827&18785\\
		\hline
		$	\pi^{+}\pi^{-}$&160&19693&61491&1490&572&929&0&7018&59&30890&9646&131948\\
		\hline
		$	\geq 3\pi$ &794&11554&16293&2845&5724&6810&01&30554&1907&14505&190602&281589\\
		\hline
		total&421970&278233&764584&8404&28856&38212&01&161235&2876&61226&234403&2000000\\
		\hline	
	\end{tabular}
	\caption{Occupancy of primary and final state hadronic systems for 2 Million events in GENIE(v3.00.06) using DUNE flux on $^{40}$Ar target for $\nu_{\mu}$-nucleus CC interactions, without applying detector cuts. Different topological groups for primary and final state systems were made on the basis of number of pions produced event wise.}
	\label{table1}
\end{table}

\begin{table}[h!]
	\centering
	\begin{tabular}{|c|c|c|c|c|c|c|c|c|c|c|c|c|c| }
		\hline
		& \multicolumn{12}{|c|}{Primary Hadronic States }\\
		\hline
		Final states  & $0\pi $&$ \pi^{0} $& $\pi^{+} $&$ \pi^{-} $&$ 2\pi^{0}  $&$ 2\pi^{+} $&$ 2\pi^{-} $&$ \pi^{0} \pi^{+}$ &$ \pi^{0}\pi^{-} $&$ \pi^{+}\pi^{-}$&$ \geq 3\pi$ & total\\
		\hline
		$		0\pi $&433572 &31896 &105429 &1293 &258 &315 &0 &2898 &17 &1388 &155 &577221\\
		\hline
		$	\pi^{0}$ & 6241&165238&28650&318&3433&166&0&21266&285&905&1434&227936\\	
		\hline	
		$	\pi^{+}$&9768&7182&405364&49&144&3478&0&20052&03&10336&1121&457497\\
		\hline	
		$	\pi^{-}$&4342&8160&6090&13807&165&37&01&957&223&10592&787&45161\\
		\hline
		$	2\pi^{0}$&884&3522&1818&09&14327&30&0&6834&74&226&5305&33029\\
		\hline
		$	2\pi^{+}$&119&248&4596&0&18&12374&0&4170&0&638&2803&24966\\
		\hline
		$	2\pi^{-}$&69&245&214&55&17&02&13&95&54&655&304&1723\\
		\hline
		$	\pi^{0} \pi^{+}$&615&3864&8448&04&745&978&0&163227&16&3545&10181&191623\\
		\hline
		$	\pi^{0}\pi^{-}$&436&2772&1524&117&847&21&0&1859&2970&3563&6081&20190\\
		\hline
		$	\pi^{+}\pi^{-}$&651&2351&9279&167&71&275&0&5945&45&94266&9794&122844\\
		\hline
		$	\geq 3\pi$ &423&3120&5697&73&1522&1106&01&17519&263&9403&258684&297810\\
		\hline
		total&457120&228598&577108&15892&21547&18782&15&244822&3950&135517&296649&2000000\\
		\hline	
	\end{tabular}
	\caption{Occupancy of primary and final state hadronic systems for 2 Million events in NuWro(v19.02.2) using DUNE flux on $^{40}$Ar target for $\nu_{\mu}$-nucleus CC interactions, without applying detector cuts. Different topological groups for primary and final state systems were made on the basis of number of pions produced event wise.}
	\label{table2}
\end{table}

\begin{figure}[h!]
	\centering     
	\includegraphics[scale=0.35]{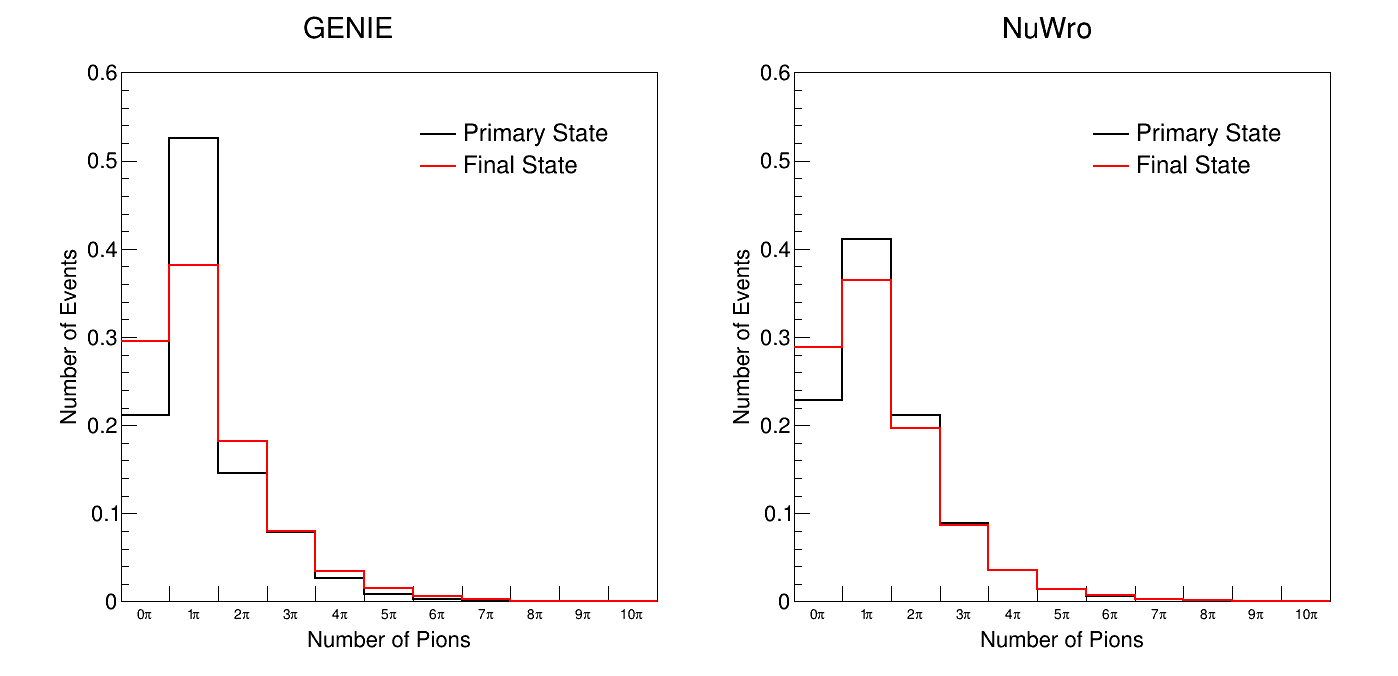}
	\caption{Figure showing number of pions produced on event by event basis in  primary and final states for GENIE (left panel) and NuWro (right panel) generators without applying detector cuts. Black lines represent primary states and red lines represent final states in both the panels.}
	\label{Combined}
\end{figure}

For GENIE, the results of simulation for 2 Million events using default values of axial masses: $M_{A}^{QE}$ =0.99 GeV/$c^{2}$, $M_{A}^{RES}$ =1.12 GeV/$c^{2}$, $M_{A}^{COH}$ =1.00 GeV/$c^{2}$, are shown in Table. \ref{table1}. The table shows the occupancy of primary and final state topologies. A comparison plot for primary and final state pions is shown in Fig. \ref{Combined} (left panel).\\

For NuWro, the results of simulation for 2 Million events using  values of axial masses: $M_{A}^{QE}$ =0.99 GeV/$c^{2}$, $M_{A}^{\pi}$ =0.94 GeV/$c^{2}$, $M_{A}^{COH}$ =1.00 GeV/$c^{2}$ and NuWro's intranuclear cascade model, are shown in Table \ref{table2}. A comparison plot for primary and final state pions is shown in Fig. \ref{Combined} (right panel).\\ 

From Table \ref{table1} (GENIE) and Table \ref{table2} (NuWro), it is clear that for two generators, in many cases, there are some large differences in both primary and final state topologies (i.e. the number of pions observed in primary and final states). The differences can also be observed from Fig.  \ref{Combined}, wherein plots for number of pions observed on event by event basis, in primary and final states have been shown for each generator. In many cases, these differences are above statistical fluctuations. For example, number of $\pi^{+}$ observed in the final state corresponding to $\pi^{0}$ in the primary state is 13618 for GENIE while the corresponding number for NuWro is 7182, the difference is more than 47 $\%$. The number of $\pi^{0}$ observed in the final state corresponding to $\pi^{+}$ in the primary state is 78184 for GENIE while the corresponding number is 28650 for NuWro, the difference is even larger. The number of $\pi^{-}$ observed in the final state corresponding to $\pi^{-}$ in the primary state is 1787 for GENIE while the corresponding number is 13807 for NuWro. These differences are attributed to the fact that DUNE flux peaks around 2.5 GeV and in this energy region QE, RES and DIS processes all contribute significantly towards total cross-section. Although the models used to describe these processes separately are often common to a generator, still there are many differences in the way in which a particular generator takes into consideration the merging of the relative contributions in this energy region. This effect in combination with other input parameters can lead to observed differences from the same set of models.\\

Another observation from the two tables is that both the generators have larger number of zero pion (0$\pi$) topologies in the final states than in the primary states. This shows that pions are more likely to be absorbed than created during their intranuclear transport. Generally, QE processes give rise to topologies with 0$\pi$ in the primary and final states while RES and DIS processes are likely to result in events with pions in the primary and final states. Also both the generators have almost similar 0$\pi$ topologies. The differences may arise due to the difference in the values of form-factor parameters and nuclear models used in each generator.\\

Using Tables \ref{table1} and \ref{table2}, percentage of events without pion $(0\pi)$, with exactly one pion (1$\pi$) and with more than one pion ($>$1$\pi$) is shown in Table \ref{table3}. It is observed that single pion production (1$\pi$)  is favoured in GENIE while multiple pion production is more in case of NuWro.

\begin{table}[h!]
	\centering
	\begin{tabular}{|c|c|c|}
		\hline
		Pions & GENIE & NuWro\\
		\hline
		$	0\pi$ & 21.1$\%$ (29.5$\%$) &22.9$\%$ (28.9$\%$) \\
		\hline
		$	1\pi^{0}$ & 13.9$\%$ (13$\%$) &11.4$\%$ (11.4$\%$) \\
		\hline
		$	1\pi^{+}$ & 38.2$\%$ (23.7$\%$) &28.9$\%$ (22.9$\%$) \\
		\hline
		$	1\pi^{-}$ & 0.4$\%$ (1.4$\%$) &0.8$\%$ (2.3$\%$) \\
		\hline
		$	1\pi$ & 52.6$\%$ (38.2$\%$) &41.1$\%$ (36.5$\%$) \\
		\hline
		$	>$1$\pi $& 26.3$\%$ (32.3$\%$)&36.1$\%$ (34.6$\%$) \\
		\hline
	\end{tabular}
	\caption{Percentage of events without pion (0$\pi$), with exactly one pion (1$\pi$) and with more than one pion ($>$1$\pi$), at primary vertex. Values in brackets refer to results after final state interactions.}
	\label{table3}
\end{table}

\begin{table}[h!]
	\centering
	
	\begin{tabular}{|c|c|c|}
		\hline
		Process & GENIE & NuWro\\
		\hline
		$	\pi^{0} \longrightarrow \pi^{0}$ & 54.6$\%$ &72.3$\%$ \\
		\hline
		$	\pi^{+} \longrightarrow \pi^{+} $& 55.5$\%$ &70.2$\%$ \\
		\hline
		$	\pi^{0} \longrightarrow 0\pi's $& 15.8$\%$ &14$\%$ \\
		\hline
		$	\pi^{+} \longrightarrow 0\pi's $& 17.4$\%$ &18.3$\%$ \\
		\hline
		$	\pi^{0} \longrightarrow \pi^{+}$ & 4.9$\%$ &3.1$\%$ \\
		\hline
		$	\pi^{0} \longrightarrow \pi^{-} $& 4.7$\%$ &3.6$\%$ \\
		\hline
		$	\pi^{+} \longrightarrow \pi^{0} $& 10.2$\%$ &5$\%$ \\
		\hline	
	\end{tabular}
	\caption{Rate of events with single pion or no pion in the final state if there was single pion in the primary state, without applying detector cuts.}
	\label{table4}
\end{table}

\begin{figure}[h!]
	\centering     
	\includegraphics[scale=0.35]{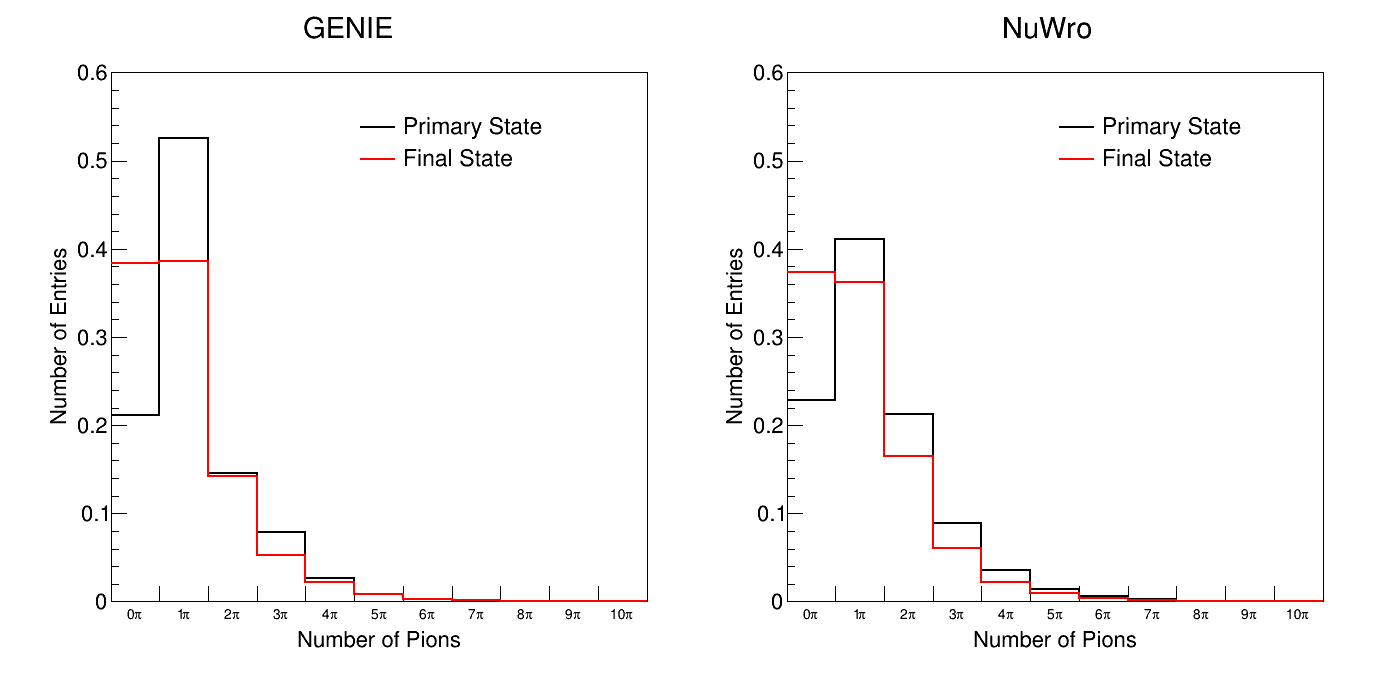}
	
	\caption{Figure showing number of pions produced on event by event basis in  primary and final states for GENIE (left panel) and NuWro (right panel) generators with KE detector threshold cuts. Black lines represent primary states and red lines represent final states in both the panels.}
	\label{Combined_cuts}
\end{figure}

Tables \ref{table1} and \ref{table2} contain important information about final state interactions. To extract this information, we have compiled a summary table (Table \ref{table4}). The summary table shows directly the topology changing effect due to intranuclear hadron transport. In this table, we show for each generator, out of all events with a given primary state topology, the fraction of those which have both the primary and final state topologies. The first two rows show the transparency of the nucleus. These rows give us the percentage of events with a single pion in the primary state that will still have the single pion in the final state. We observe that it is more likely that the pion created in the primary vertex will not re-interact. The next rows show the fraction of pions absorbed and the remaining three rows show the effect due to charge exchange. It is clear that current version of NuWro (v19.02.2) has high transparency than current version of GENIE (v3.00.06). This may be due to response of a generator to absorption and charge exchange processes (NuWro could have too little and GENIE could have too much). Taking into consideration the complex nature of final state  interactions, one can say that the agreement is still good despite these differences. The analysed MC generators give quite a similar results. As single pion events form the main background in neutrino oscillation experiments, the analysis done may be quite useful.\\

CC1$\pi^{+}$ to CCQE cross-section ratio can be calculated from the corresponding numbers in our simulation for both the generators. Table \ref{table5} shows comparison of ratio for two generators. For primary hadronic sates, GENIE ratio is higher than that of NuWro which shows that GENIE has higher cross-section for single $\pi^+$ production. The corresponding ratio for MiniBooNE data rescaled for isoscalar target and corrected for FSI's is (1.318$\pm$ 0.247) for neutrino energy of (2.1 $\pm$ 0.3) GeV\cite{MiniBooNE:2009koj}. Also percentage change in ratio due to FSI is 56 for GENIE and 39 for NuWro which indicates that GENIE shows more effects of FSI on $\pi^{+}$ than NuWro.\\

\begin{table}[h!]
	\centering
	\begin{tabular}{|c|c|c|c| }
		\hline
		Generator &$ 1\pi^{+}/0\pi$ ( Primary state total) & $1\pi^{+}/0\pi$ ( Final state total) & Percentage change after FSI\\
		\hline
		GENIE & 1.8 &0.8 & 56 \\
		\hline
		NuWro & 1.3 &0.8 & 39 \\
		\hline
		
	\end{tabular}
	\caption{The ratio 1$\pi^{+}/0\pi$ for GENIE (v3.00.06) and NuWro (v19.02.2) using DUNE flux and $^{40}$Ar target}
	\label{table5}
\end{table}

The analysis done and results shown up to now were with 100 percent detector resolution (no detector cuts applied). We will now show the results with  detector cuts on kinetic energy (KE) thresholds taken from DUNE CDR \cite{DUNE:2016ymp, DUNE:2016rla, DUNE:2015lol}). The KE thresholds used in our analysis are summarized in Table \ref{table6}. These thresholds are applied to the outgoing pions and only pions with KE above these thresholds will be identified in the final state. The results of simulation for different topologies of pions observed in the primary state and final states, with KE detector threshold cuts, are shown in Tables \ref{table7} and \ref{table8}. Again there are differences in both primary and final state topologies more or less similar to differences observed in Tables \ref{table1} and \ref{table2}. The differences can also be observed from Fig. \ref{Combined_cuts} where number of pions observed in primary and final states are plotted on event by event basis. \\

\begin{table}[h!]
\centering
\begin{tabular}{|c|c| }
	\hline
	Particle type & KE detection threshold\\
	\hline
	$e^{\pm}$, $\mu^{\pm}$, $\gamma$ & 30MeV\\
	\hline
	$\pi^{\pm}$ & 100MeV \\
	\hline
	p, n, other & 50MeV\\
	\hline
	
\end{tabular}
\caption{Detection thresholds for various final state particles at DUNE}
\label{table6}
\end{table}

If a pion has KE below the detector threshold, the pion is not identified in the final state. Thus, in a final state with single pion topology, if the KE of the pion is below the detector threshold, the pion will not be detected in the final state and the event becomes a fake event for 0$\pi$ topology thereby increasing its number. This is clear from Tables \ref{table1}, \ref{table2}, \ref{table7} and \ref{table8} that both the generators have fairly larger number of events (entries) in final state for 0$\pi$ topology when KE threshold cuts have been applied than without such cuts. For GENIE, the total number of 0$\pi$ events in the final state without  detector cuts (Table \ref{table1}) is 590212 and with detector cuts the number is 766776 (Table \ref{table7}) while for NuWro the corresponding numbers are 577221 (Table \ref{table2}) and 746450 (Table \ref{table8}) respectively. The change is about 23$\%$ in for both the generators. On the other hand, in an event with multiple pions in the final state, if one pion has KE below the detector threshold, that pion will not be identified and the topology of final state changes thereby increasing the number of events for the resulting topology state. Using Tables \ref{table7} and \ref{table8}, percentage of events in final state without pion $(0\pi)$, with exactly one pion (1$\pi$) and with more than one pion ($>$1$\pi$), with detector cuts, is shown in Table \ref{table9}. For comparison, the corresponding results without applying detector cuts are also given in the same table (i.e. Table \ref{table9}). Thus, it is clear that detector thresholds can give significant changes in pion topologies in the final state and should be taken into account for correct analysis of pions produced at the primary vertex.\\

  \begin{table}[h!]
  	\centering
  	\begin{tabular}{|c|c|c|c|c|c|c|c|c|c|c|c|c|c| }
  		\hline
  		& \multicolumn{12}{|c|}{Primary Hadronic States }\\
  		\hline
  		Final states  & $0\pi $&$ \pi^{0} $& $\pi^{+} $&$ \pi^{-} $&$ 2\pi^{0}  $&$ 2\pi^{+} $&$ 2\pi^{-} $&$ \pi^{0} \pi^{+}$ &$ \pi^{0}\pi^{-} $&$ \pi^{+}\pi^{-}$&$ \geq 3\pi$ & total\\
  		\hline
  		$	0\pi$ & 412822 & 80693&247916&1472&1060&3365&0&9361&197&2189&7701&766776\\
  		\hline
  		$	\pi^{0}$& 2898 &135349&66084&519&9203&1398&0&28002&177&1085&13053&257768\\	
  		\hline	
  		$	\pi^{+}$&4218&16922&371771&791&444&12399&0&27860&27&9658&12145&456235\\
  		\hline	
  		$	\pi^{-}$&891&15670&19761&1757&493&436&0&2550&163&9631&6086&57438\\
  		\hline
  		$	2\pi^{0}$&38&3696&5373&98&11178&266&0&5337&67&320&12437&38810\\
  		\hline
  		$	2\pi^{+}$&34&632&3379&89&61&12840&0&3210&07&820&8806&29878\\
  		\hline
  		$	2\pi^{-}$&04&199&431&34&63&24&0&183&22&475&990&2425\\
  		\hline
  		$	\pi^{0} \pi^{+}$&242&6301&11874&253&1365&2557&0&59936&94&2118&19201&103941\\
  		\hline
  		$	\pi^{0}\pi^{-}$&42&3123&2441&206&1346&153&0&2220&375&2032&8017&19955\\
  		\hline
  		$	\pi^{+}\pi^{-}$&141&8662&28289&1011&563&1305&01&6528&65&24269&18040&88874\\
  		\hline
  		$	\geq 3\pi$ &640&6986&7265&2174&3080&3469&0&16048&1682&8629&127927&177900\\
  		\hline
  		total&421970&278233&764584&8404&28856&38212&01&161235&2876&61226&234403&2000000\\
  		\hline	
  	\end{tabular}
  	\caption{Occupancy of primary and final state hadronic systems for 2 Million events in GENIE(v3.00.06) using DUNE flux on $^{40}$Ar target for $\nu_{\mu}$-nucleus CC interactions with KE threshold detector cuts. Different topological groups for primary and final state systems were made on the basis of number of pions produced event wise.}
  	\label{table7}
  \end{table}
  
  \begin{table}[h!]
  	\centering
  	\begin{tabular}{|c|c|c|c|c|c|c|c|c|c|c|c|c|c| }
  		\hline
  		& \multicolumn{12}{|c|}{Primary Hadronic States }\\
  		\hline
  		Final states  & $0\pi $&$ \pi^{0} $& $\pi^{+} $&$ \pi^{-} $&$ 2\pi^{0}  $&$ 2\pi^{+} $&$ 2\pi^{-} $&$ \pi^{0} \pi^{+}$ &$ \pi^{0}\pi^{-} $&$ \pi^{+}\pi^{-}$&$ \geq 3\pi$ & total\\
  		\hline
  		$		0\pi $&446462 &66167 &193147 &7180 &1828 &1751 &01 &16182 &216 &8463 &5053 &746450\\
  		\hline
  		$	\pi^{0}$ &3327 &145671 &20119 &161&7584&252&0&42866&1303&1679&11212&234474\\	
  		\hline	
  		$	\pi^{+}$&4487&5327&347435&50&199&6777&0&42241&22&23051&8966&438555\\
  		\hline	
  		$	\pi^{-}$&2129&5714&4835&8336&217&74&04&1605&435&22155&5525&51029\\
  		\hline
  		$	2\pi^{0}$&221&1519&751&04&10210&15&0&4642&44&170&9744&27320\\
  		\hline
  		$	2\pi^{+}$&20&112&1781&0&09&8775&0&2690&0&598&6839&20824\\
  		\hline
  		$	2\pi^{-}$&15&80&71&24&09&0&10&66&32&576&592&1475\\
  		\hline
  		$	\pi^{0} \pi^{+}$&144&1486&3160&01&443&592&0&123904&15&2374&24910&157031\\
  		\hline
  		$	\pi^{0}\pi^{-}$&114&995&530&51&516&13&0&1419&1767&2317&13086&20808\\
  		\hline
  		$	\pi^{+}\pi^{-}$&155&747&3536&68&42&199&0&3584&36&70881&22185&101433\\
  		\hline
  		$	\geq 3\pi$ &46&780&1443&17&490&334&0&5621&80&3253&188537&200601\\
  		\hline
  		total&457120&228598&577108&15892&21547&18782&15&244822&3950&135517&296649&2000000\\
  		\hline	
  	\end{tabular}
  	\caption{Occupancy of primary and final state hadronic systems for 2 Million events in NuWro(v19.02.2) using DUNE flux on $^{40}$Ar target for $\nu_{\mu}$-nucleus CC interactions with KE detector threshold cuts. Different topological groups for primary and final state systems were made on the basis of number of pions produced event wise.}
  	\label{table8}
  \end{table}

    \begin{table}[h!]
	\centering
	\begin{tabular}{|c|c|c|c|c|}
		\hline
		& \multicolumn{2}{|c|}{With Detector Cuts} & \multicolumn{2}{|c|}{Without Detector Cuts}\\
		\hline
		Pions & GENIE & NuWro &GENIE &NuWro\\
		\hline
		$	0\pi$ & 38.3$\%$  &37.3$\%$& 29.5$\%$ &28.9$\%$  \\
		\hline
		$	1\pi^{0}$ & 12.9$\%$  &11.7$\%$& 13$\%$ &11.4$\%$ \\
		\hline
		$	1\pi^{+}$ & 22.8$\%$  &21.9$\%$ & 23.7$\%$ &22.9$\%$ \\
		\hline
		$	1\pi^{-}$ & 2.9$\%$  &2.6$\%$ & 1.4$\%$ &2.3$\%$ \\
		\hline
		$	1\pi$ & 38.6$\%$  &36.2$\%$ & 38.2$\%$ &36.5$\%$ \\
		\hline
		$	>$1$\pi $& 23.1$\%$ &26.5$\%$  & 32.3$\%$ &34.6$\%$ \\
		\hline
	\end{tabular}
	\caption{Percentage of events in final state without pion (0$\pi$), with exactly one  pion (1$\pi$) and with more than one pion ($>$1$\pi$), with KE detector threshold cuts. The results obtained without applying these cuts are also given  for comparison.}
	\label{table9}
\end{table}

\begin{table}[h!]
	\centering	
	\begin{tabular}{|c|c|c|c|c|}
		\hline
		& \multicolumn{2}{|c|}{With Detector Cuts} & \multicolumn{2}{|c|}{Without Detector Cuts}\\
		\hline
		Process & GENIE  & NuWro  & GENIE & NuWro\\
		\hline
		$	\pi^{0} \longrightarrow \pi^{0}$ & 48.6$\%$  &63.7$\%$  &54.6$\%$ &72$.3\%$ \\
		\hline
		$	\pi^{+} \longrightarrow \pi^{+} $& 48.6$\%$ &60.2$\%$ &55.5$\%$ &70.2$\%$\\
		\hline
		$	\pi^{0} \longrightarrow 0\pi's $& 29$\%$ &28.9$\%$  &15.8$\%$ &14$\%$\\
		\hline
		$	\pi^{+} \longrightarrow 0\pi's $& 32.4$\%$ &33.5$\%$ &17.4$\%$ &18.3$\%$ \\
		\hline
		$	\pi^{0} \longrightarrow \pi^{+}$ & 6.1$\%$ &2.3$\%$  &4.9$\%$ &3.1$\%$\\
		\hline
		$	\pi^{0} \longrightarrow \pi^{-} $& 5.6$\%$ &2.5$\%$ &4.7$\%$ &3.6$\%$\\
		\hline
		$	\pi^{+} \longrightarrow \pi^{0} $& 8.6$\%$ &3.5$\%$ &10.2$\%$ &5$\%$\\
		\hline	
	\end{tabular}
	\caption{Rate of events with single pion or no pion in the final state if there was single pion in the primary state with KE detector threshold cuts. The results obtained without applying these detector cuts are also shown for comparison.}
	\label{table10}
\end{table}

The summary of results obtained from Tables \ref{table7} and \ref{table8} is shown in Table \ref{table10}. It is clear from Tables \ref{table4} and \ref{table10} that applied threshold cuts have shown certain differences for all processes and appreciable differences for some of the processes. In the process, $\pi^{0}$ $\longrightarrow$ $\pi^{0}$, the percentage of event rate has been reduced from 55 (without detector cuts) to 49.1 (with detector cuts) in GENIE and from 71.5 (with detector cuts) to 62.8 (without detector cuts) in NuWro. Similar trend is followed in the process $\pi^{+}$ $\longrightarrow$ $\pi^{+}$. Third and fourth rows show that percentage of pion absorbed is more in the presence of KE detector threshold cuts which is obvious as pions with KE less than detector threshold energy are not taken in the final state and this decreases the percentage of final state pions which in a sense is equivalent to a pion absorbed and not reaching the final state thereby increasing the percentage of pion absorbed. The last three rows show the effect of charge exchange where there is not much difference for the processes with and without detector threshold cuts. As is clear from Table \ref{table10}, the two generators are following similar trends for results with and without detector threshold cuts. The results are also shown in Fig. \ref{Combined_cuts} where plots for number of pions observed on event by event basis in primary and final states, with KE detector threshold  cuts, have been shown for each generator. It is clear that single pion events which come from RES scattering are least effected by KE detector threshold  cuts.\\

\section{Summary and Conclusions}
In the present work, we report an extensive analysis of effect of final state interactions on pion production for $\nu_{\mu}$-nucleus interactions on $^{40}$Ar target. The results are obtained for 100$\%$ detector resolution and then compared with the results obtained after applying KE  detector threshold cuts. The latest versions of GENIE and NuWro generators were used as simulation tools. We have shown in Tables \ref{table1} and \ref{table2} that two generators, despite having similar sets of models, can give some differences in results of pion production both in primary and final states. This may be due to differences in the implementation of models and other input parameters used in two generators.\\
Both the generators show almost similar effects of final state interactions on pions during their intranuclear transport after being produced at primary vertex. This is clear from Table \ref{table3}. It is clear from Fig. \ref{Combined} that final state interactions create a  difference between pions observed in the detector (final state pions) and pions produced at the primary vertex (primary state pions).\\
Tables \ref{table1} and \ref{table2} also make it clear that both the generators have larger number of 0$\pi$ topologies in the final state than initial state leading to the conclusion that pions are more likely to be absorbed than created during their intranuclear transport.\\
We have also explained the topology changing effect during intranuclear hadron transport, using summary table (Table \ref{table4}). The effect is more for GENIE than NuWro which shows that current version of NuWro shows high transparency than the current version of GENIE.\\
CC1$\pi^{+}$ to CCQE cross-section ratio is calculated from the corresponding numbers in our simulation for both the generators. Table \ref{table5}, shows that GENIE has a higher cross-section for single $\pi^{+}$ production than NuWro at the primary vertex. Also percentage change in ratio after FSI is higher for GENIE than NuWro. On the whole, taking into consideration the complex nature of FSI, two generators are giving quite a similar results.\\
Finally, we applied KE threshold cuts on the outgoing particles which are to be detected by the detector. Significant differences are observed on the final state particles with the application of detector threshold cuts. This can be seen from Table \ref{table10} and Figs. \ref{Combined} and \ref{Combined_cuts}. Thus detector threshold are to be taken into account to get correct information  about the particles produced in neutrino interactions and there is need to improve the detector technology to improve the detector threshold for better results.\\

Our results indicate that for a neutrino oscillation experiment like DUNE, the best strategy should be to have authentic accuracy of nuclear models used in neutrino event generators applied for simulation and for that a canonical neutrino event generator is required. As it is critical to understand nuclear effects, a clear understanding of hadronic physics of neutrino-nucleus interactions is required. The detector thresholds are to be minimized by using advanced detector technology.\\

In future, more neutrino event generators would be used to compare the results. Also 2p2h events (also called meson exchange(MEC) events) which have signatures indistinguishable from QE events and act as background for QE events, would be taken into consideration in our future simulation work.\\

\end{document}